\documentclass[journal]{IEEEtran}
\usepackage{graphicx}
\usepackage{cite}
\usepackage{makecell}
\usepackage{multirow}
\usepackage{booktabs}
\usepackage{array}
\usepackage{amssymb}

\ifCLASSINFOpdf
\else
\fi
\begin{document}
%
\title{The Internet of People: A Survey and Tutorial}
%
%
%

\author{Feifei~Shi,
        Wenxi~Wang,
        Hang~Wang,
        and~Huansheng~Ning,~\IEEEmembership{Senior Member,~IEEE}
\thanks{F. Shi, W. Wang, H, Wang, and H. Ning are with School of Computer and Communication Engineering, University of Science and Technology Beijing, Beijing, 100083, China. Corresponding email: ninghuansheng@ustb.edu.cn}
\thanks{H. Ning is also with Beijing Engineering Research Center for Cyberspace Data Analysis and Applications, Beijing, 100083, China.}}

\maketitle

\begin{abstract}
Along with the rapid development of Internet and cyber techniques, the Internet of People (IoP) has been emerging as a novel paradigm which establishes ubiquitous connections between social space and cyberspace. Unlike the Internet of Things (IoT), human nodes in IoP are not only terminal users, but also significant participants bonded with tighter relationships. Note that IoP has deeply influenced our lives. For example, the prosperity of social media enables us to build and maintain new relationships without considering physical boundaries. In this paper, we give a comprehensive overview of IoP by comparing it with IoT, introduce its enabling techniques from aspects of sensing, communication and application. In addition, we summarize and compare popular IoP-enabled platforms according to different functions, and finally envision open issues needed to be further discussed in IoP. It provides constructive suggestions, lays solid foundations, and makes a basic blueprint for the future development of IoP paradigm.
\end{abstract}


%
\IEEEpeerreviewmaketitle

\section{Introduction}
%
%
%
%
\IEEEPARstart{O}{ver} the past years, the Internet of Things (IoT) has been penetrating into every corner of our daily life, ranging from living homes, intelligent transportation to smart healthcare and finance \cite{A1}. IoT is such a paradigm that concentrates on the seamless connections between objects and things, and benefits a lot from the deep convergence between cyber and physical spaces \cite{A2}. Meantime, due to the fast proliferation of smart sensors and personal intelligent devices, users are gradually bonded together with tighter relationships. Their social identities and relationships are also being mapped into cyberspace, and the interconnections between cyber and social spaces are getting increasingly closer. A novel paradigm around humans, the so-called Internet of People (IoP) is evolving toward a mature future \cite{A3}.

The IoP, as name implies, describes an interconnected network composed of various human nodes. It refers to the digitalization between humans and their generated information and data, where various personal connected, intelligent devices that could maintain social interconnections in cyberspace \cite{A4}. In other words, humans become significant participants and are able to meet, negotiate, communicate, and work with each other in  virtual cyberspace. The IoP provides more possibilities for humans to carry out social activities, establish and maintain social relationships, and exchange data and information with overcoming physical boundaries, for example, the social online network is such a representative miniature of IoP \cite{A5}.

These years with the advanced progress of cyberspace and cyber techniques, IoP has emerged at a faster speed, and is expected to become the frontier in next few years. Related researches have been advancing steadily, and in 2015, Ma and Ning initially launched the IEEE International Conference on Internet of People (IoP) \cite{A6}. Since then, a valuable research area has been opened up. In this survey, we provide an overall comprehension of IoP, scoping from its enabling techniques and typical platforms to open challenges and issues. The main contributions of this paper are as follows:

\begin{itemize}
\item Provide a comprehensive survey and tutorial on IoP, so as to provide valuable guidance for future development.
\item Introduce and overview enabling techniques of IoP, and analyze typical IoP platforms, which makes a clear blueprint for IoP paradigm.
\item Envision challenging issues faced with IoP, and point out future directions.
\end{itemize}

The remainder of this paper is arranged as follows: Section 2 mainly focuses on the difference between IoP and IoT. Section 3 introduces popular enabling technologies of IoP. Section 4 analyzes typical IoP platforms. Section 5 envisions challenging issues and points out prospects.

\section{IoT versus IoP}

In order to provide a understandable introduction of IoP, in this section, we expound the difference between IoT and IoP from the perspective of space convergence. First of all, we need to emphasize that there is absolutely no substitution relationship between IoT and IoP, but is the development direction of cyberspace and cyberspace technology. Thanks to the advances of IoT, sensors and smart devices have sprung up in large numbers, hence most people begin to be closely linked, and finally form interconnected IoP.

\begin{figure}[!hb]
\centering
\includegraphics[width=8 cm]{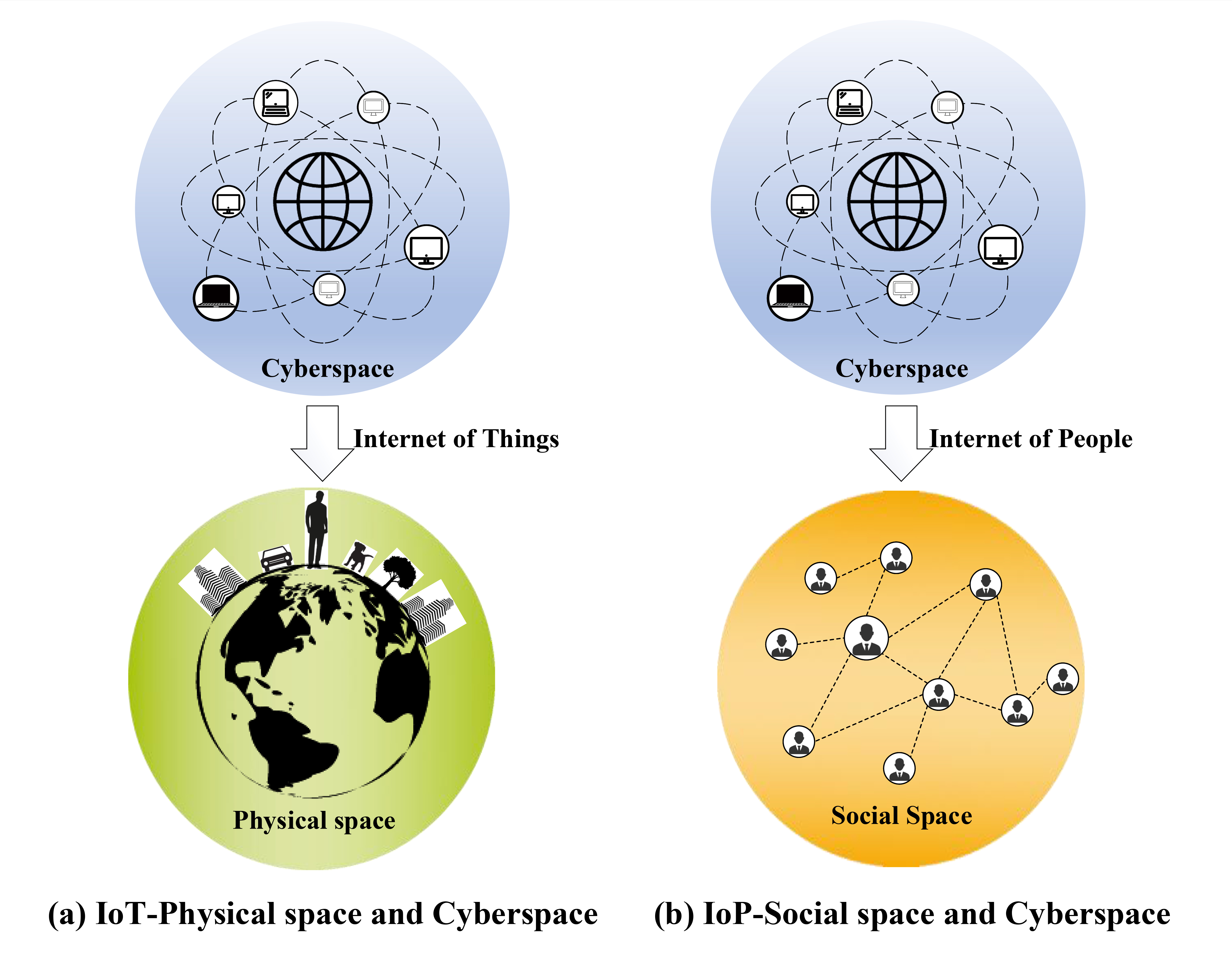}
\caption{The difference between IoT and IoP from the perspective of space convergence.}
\label{fig1}
\end{figure}

As can be seen in Figure \ref{fig1}, a general comparison between IoT and IoP is made from the perspective of space convergence. IoT refers to the deep relationships between physical space and cyberspace. That is to establish physical-based ubiquitous connections between kinds of objects and things. Usually for IoT, the three-tier architecture composed of sensing layer, network layer and application is such popular, and supports high-level services \cite{A7}.

While for IoP, it mainly represents the seamless connections between humans, and benefits a lot from the deep convergence between social space and cyberspace. In other words, it provides stronger links for people in social space that could overcome traditional physical boundaries. Due to the widespread of Internet and cyber techniques, people could own similar attributes, roles, and relationships in cyberspace, and are able to participate in various online activities, establish and maintain relationships, form virtual social communities and so forth. It is reported by Statista that there were 4.66 billion active Internet users worldwide as of January 2021 \cite{A8}. They can communicate and cooperate more easily, generate a large amount of data and information, and contain great values as well.

\section{Enabling Techniques}

As mentioned above, we define IoP as a networked paradigm enabling ubiquitous connections between social space and cyberspace, which is human-centered. Humans need to make interactions with each other, and generated, transmit, process, and analyze information and data in detail. Therefore, it is significant to understand enabling techniques developed in IoP, so as to make the most use of it.

In order to provide a comprehensive introduction of IoP enabling techniques, we refer to the three-tier architecture in IoT, and elaborate detailed technologies from the aspects of sensing, communication, and application. First, we need to declare that, humans are the most substantial elements in IoP. In addition, there will also be various intelligent sensors and personal devices that help humans be interconnected to the social networks.

\subsection{Sensing techniques}

When it comes to sensing techniques in IoT, most of them are related to environment contexts, attributes of things or objects, whose aim is to sense the dynamics of surroundings. However, for human-centric IoP, its main task is to find and gather information of human nodes. Thanks to the advances of Internet, the sensing information evolves from basic identification, physiological information, to more sophisticated information of social attributes and preferences. In other words, humans in IoP are equipped with increasingly complete information, and could more accurately map individuals in real life.

In general, the sensing techniques involved in IoP could be classified as active sensing and passive sensing. On the one hand, active sensing refers that humans in IoP provide information actively, such as filling in relevant basic information during registration, associating their own identities with unique social roles, etc. Hence, techniques such as identification modeling, user portrait modeling, semantic ontologies and so forth are so important for helping model humans in IoP \cite{A9,A10,A11}. In addition, humans have the rights to search and collect information they are interested or need, for instance, they may need to establish relationships with someone in the same online community, therefore they could get specified recommendations based on search demands.

On the other hand, passive sensing is also a common kind technology. It represents that humans do not need to collect information actively, while the IoP network could recommend potential information based on your past history, preferences, and habits etc. In this case, algorithms such as user portrait modeling, accurate matching, and intelligent recommendation are commonly. For instance, Ning proposed and demonstrated a friend recommendation system based on big-five personality traits and hybrid filtering, which could recommend friend information based on respective personality \cite{A12}.

\subsection{Communication techniques}

As we all know, IoT pays more attention to data exchange and information sharing between things themselves, while in IoP, the main elements are humans, and we will concentrate more on how human nodes are linked together, and how to communicate with each other.

In traditional social space, the establishment of human relationships depends on geographic location, family history, and community activities etc. In a span of a decade, the connections have been strengthened and enabled by Internet and cyber techniques, and most rely on intelligent sensors, personal devices, and mobile communication tools to maintain communications. In other words, it is a kind of device-supported communication. In this section, we analyze typical wired and wireless communications that would be widely adopted in IoP.

\subsubsection{Wired communications}

Usually wired communications refer to transmit data or information via tangible media, such as metal wires, wired cables, and optical fibers. On the one hand, the communication is much more stable and reliable, and suffer from little disturbance from the dynamic environments. However, for wired communication, the biggest disadvantage is the poor flexibility which must rely on large cables or infrastructure. Therefore, it is mostly widely used in areas of industrial manufacture, where there are many large equipments that need to be connected with cables.

In IoP, since humans are mostly interconnected with intelligent personal sensors, devices or computers, the most common wired communications depend on these among sensors, devices, or computers. For example, wired cables could be used to connect personal computers within a certain range in a LAN, so as to find new friends or establish new contacts. In addition, there is also some other wired ways for communications, such as Serial Communications, Thunderbolt and USB interfaces that are generally used when connecting external devices.

\subsubsection{Wireless communications}

\begin{table*}[!ht]
\newcommand{\tabincell}[2]{\begin{tabular}{@{}#1@{}}#2\end{tabular}}
\centering
\caption{The comparison between five common wireless communication techniques.}
\label{Tab1}
\begin{tabular}{cccc}
\toprule
Name & Features & Protocol & Physical Range \\
\midrule
Bluetooth & \multicolumn{1}{m{6cm}}{A short-range wireless communication enabling devices to communicate directly with each other. It is deeply influenced by required distance, and is usually adopted among smart phones, watches and headphones existing in a relatively small space.} & IEEE 802.15.1 & \multicolumn{1}{m{6cm}}{Typically less than 10 m. Taking Bluetooth 5 as an example, its range is up to 400 m.} \\
\specialrule{0em}{3pt}{3pt}
Wi-Fi & \multicolumn{1}{m{6cm}}{A wireless ``Ethernet'' in local network area. It adopts an asymmetric communication mode. That is to say, all data and information need to be transmitted through a wireless center. It could be used in smart homes, factories and public offices.} & IEEE 802.11 & \multicolumn{1}{m{6cm}}{Generally with a scope between 100 and 300 m.} \\
\specialrule{0em}{3pt}{3pt}
Zigbee & \multicolumn{1}{m{6cm}}{Zigbee communication mainly depends on low-power digital radios, and is widely adopted in home automation, small-scale data collection and so forth. It is a low-power, low-cost wireless ad hoc network.} & IEEE 802.15.4 & \multicolumn{1}{m{6cm}}{Usually with a scope of 10-100 m, while it is also impacted with power output and environment contexts.} \\
\specialrule{0em}{3pt}{3pt}
5G Cellular & \multicolumn{1}{m{6cm}}{5G cellular is composed of cells where devices could connect with Internet or telephone network via radio waves. In 5G cellular networks, the bandwidth and download speed are greatly improved.} & 5G protocols & \multicolumn{1}{m{6cm}}{The coverage radius of 5G base stations is generally about 100 to 300 m.} \\
\specialrule{0em}{3pt}{3pt}
Li-Fi & \multicolumn{1}{m{6cm}}{Instead of depending on radio waves for data transmission, Li-Fi is kind of light-based Wi-Fi that could transmit data with ultraviolet, infrared and visible light. This wireless communication would be well applied in areas susceptible to electromagnetic interference like hospitals.} & Li-Fi wireless protocol & \multicolumn{1}{m{6cm}}{In open areas, Li-Fi's coverage is up to 10 m.}\\
\bottomrule
\end{tabular}
\end{table*}

Compared with wired communications, wireless communications depend on  radio, electromagnetic wave and other ways  for long-distance communication, instead of external media. Therefore, wireless communications could get rid of the shackles of cables, and has advantages of efficient installation, convenient maintenance, and strong capacity expansion. Nowadays, there have been some researches focusing on wireless communications, for example, Zou gives an overview of technical challenges, recent advances, and future trends of wireless transmissions \cite{A13}. As can be seen in Table \ref{Tab1}, we compare five common wireless networking technologies, focusing its main features, communication protocols and range scope. Here, we take Bluetooth and Wi-Fi as representative examples.

\begin{itemize}
\item Bluetooth: Bluetooth is a short-range communication standard usually used among mobile devices, and managed by the Bluetooth Special Interest Group (SIG) with more than 35,000 members \cite{A14}. With its popularization, Bluetooth has been widely employed in various smart products, such as mobile phones, personal laptops, intelligent watches, as well as headphones designed by Apple and Google etc. With Bluetooth standard, it is easier to link humans together, while the distance is typically limited with the maximum of approximately 30 feet. In Bluetooth 5, the range scope is up to 400 m. Whatever, the IoP communication with Bluetooth must be within predefined distance.
\item Wi-Fi: Wi-Fi is composed of series of network protocols according to the IEEE 802.11, and widely used in local areas such as smart homes, factories and offices. Compared with point-to-point communications of Bluetooth, Wi-Fi is access point-centered and terminal devices connect with specified router centers, so as to share and exchange information. Generally speaking, the wireless routers have a certain range of coverage between 100 and 300 m, which are also influenced by external factors such as buildings, objects, and other barriers of signal block. Hence, for Wi-Fi communications in IoP, the closer to the routers or network cards, the stronger the signal and the better the quality.
\end{itemize}

\subsection{Application techniques}

Compared with application layer in IoT, we propose that the ultimate aim for IoP is to provide human-centric applications and services, that could intelligently satisfy personalized requirements. Upon this, we introduce three basic application techniques in this section, including Social Network Analysis (SNA), social computing, and intelligent service. SNA refers to analyze the structures, humans and edge relationships with graph theories, so as to provide insight into social influences. Social computing mainly represents exploring social behavior depending on computational techniques. At last, techniques related to human-centric service such as user portrait modeling and intelligent recommendation are so substantial, therefore to provide with the most appropriate services.

\subsubsection{Social Network Analysis (SNA)}

As mentioned above, SNA could be regarded as an interdisciplinary analysis from aspects of both sociology and information science. It could be traced back to 1980s when there were already some solid researches around SNA. For example, Scott made an overall introduction of SNA developments, origins, models, and methods \cite{A17,A18}. Generally speaking, there are some metrics used in SNA in order to better understand how humans, objects, and things interact with each other. Degree Centrality, as the most direct measure describing node centrality in network analysis, means the number of friends directly connected to the given node. The higher the degree centrality is, the more important the node is in the social network. Closeness Centrality represents the distance between given node and others in the network, and a smaller closeness centrality refers to closer relationships with others. Besides, the node with high betweenness centrality values refers to its significance as one of the most important ``bridges'' among all nodes.

In addition, link prediction is so essential in SNA that it allows to analyze the relationships between different nodes. Algorithms such as Jaccard's coefficient, common neighbors, as well as Katz, PageRank, and SimRank are all valuable methods for predicting relationships between various nodes \cite{A19}.

\subsubsection{Social computing}

Social computing, the novel computing paradigm is proposed for deepening our understanding of social systems and studying how to use computing systems to prompt humans' communication and cooperation \cite{A20}. In the networked IoP, once human nodes establish relationships with others, they may generate and exchange information with others. Hence comprehensive analysis of social data and behavior is so significant.

Some researches have already being concentrating on social computing. For example, Evans \cite{A21} proposed the concept of social computing serving as interfaces between social interactions and computations. Shi \cite{A22} introduces a human-centric social computing (HCSC) model to solve the widespread concern in society and make an adequate prediction of its development trend. It must be admitted that there are some uncertain problems in social computing, and in 2018, Jin analyzed these issues and gave series of solutions \cite{A23}. By applying data analysis and computing methods, the complicated social behavior would be explicitly analyzed and mined.

\subsubsection{Intelligent service}

Compared with IoT, IoP concentrates more on humans and their relationships, and proposes much more demanding requirements for intelligent service. However, the contradictions between personalized needs and diversified services are still one of the challenging obstacles in IoP. Hence it is essential to understand everyone's needs as much as possible, match and recommend the most expected services. In this section, we mainly illustrate two kinds of related techniques for providing intelligent services.

First of all, user portrait modeling holds a significant position since it provides detailed profiles of users, including phycological attributes, social relationships, habits and preferences. All these information would help a lot in understanding and analyzing users' behavior. Skillen provides an ontological user profile model with consideration of attributes of users, temporal and environment contexts, in order to provide with an adaptive model applicable for mobile environment \cite{A24}. In addition, Hu also adopts Word2Vec to extract user profile information from search terms of a given period, with fusing of TF-IDF to improve the efficiency \cite{A25}.

Besides, algorithms related to intelligent recommendation are also essential, since IoP would require to provide with the ``right'' services to the ``right'' person at the ``right'' time \cite{A26}. DuBois highlighted the importance of trust in making recommendations among social networks, hence he proposed a new trust metric, and applied it with clustering algorithms in social networks \cite{A27}. Gurini established a three-dimensional matrix factorization composed of sentiment, volume, and objectivity extracted and generated from their social contents, to provide temporal People-to-People Recommendation on Social Networks \cite{A28}. Gradually, scholars and professions notice that it is not enough only relying on preferred interests when doing social recommendations, and therefore, Dhelim concentrates on mining user interests in signed social networks for IoP \cite{A29}. In other words, disliked or undesired information may also be worth further attention for accurate recommendations.

\section{IoP-enabled platforms}

\begin{table*}[!ht]
\normalsize
\centering
\caption{Typical IoP-enabled platforms.}
\label{Tab2}
\resizebox{\linewidth}{!}{
    \begin{tabular}{m{2cm}m{1.5cm}m{1.7cm}m{4.5cm}|m{2cm}|m{1.5cm}|m{1.5cm}|m{1.5cm}}
    \hline
    \hline
    \multirow{2}{*}{\textbf{Name}} & \multirow{2}{*}{\textbf{Country}}  & \multirow{2}{*}{\textbf{Year of issue}} & \multirow{2}{*}{\textbf{Description}} & \multicolumn{4}{c}{\textbf{Functions}}\\
    \cline{5-8}
    \multicolumn{1}{l}{~} & \multicolumn{1}{l}{~}  & \multicolumn{1}{l}{~} & \multicolumn{1}{l|}{~} & Instant messaging & Career-oriented & Multimedia search & Broadcast interaction \\
    \hline
    Tecent QQ & China & \multicolumn{1}{m{2cm}}{1999} & The initial aim is to provide with instant messaging services based on Internet. & \checkmark & & & \checkmark \\
    \hline
    LinkedIn & U.S. & \multicolumn{1}{m{2cm}}{2002} & ``One-step career development platform'' to help employees connect with unlimited opportunities. &&\checkmark & & \\
    \hline
    Myspace & U.S. & \multicolumn{1}{m{2cm}}{2003} & Social interaction platform with multiple functions. & \checkmark & & \checkmark & \\
    \hline
    Facebook & U.S. & \multicolumn{1}{m{2cm}}{2004} & A social platform to interact with friends, colleagues, classmates and people around you. & \checkmark & &\checkmark &\\
    \hline
    Flicker & Canada & \multicolumn{1}{m{2cm}}{2004} & Provide picture service, contact service, and group service. & & &\checkmark &  \\
    \hline
    Mixi & Japan & \multicolumn{1}{m{2cm}}{2004} & Japan's largest social networking site & \checkmark & & & \\
    \hline
    YouTube & U.S. & \multicolumn{1}{m{2cm}}{2005} & Platform for users to download, watch, and share movies or short films. & & &\checkmark & \\
    \hline
    Twitter & U.S. & \multicolumn{1}{m{2cm}}{2006} & Dedicated to serving public dialogue. &&&&\checkmark \\
    \hline
    Badoo & Britain & \multicolumn{1}{m{2cm}}{2006} & Realize global synchronous communication and share their life drops among online users. & & & \checkmark & \\
    \hline
    VK & Russia & \multicolumn{1}{m{2cm}}{2006} & Russia's largest social networking site. & \checkmark & & &\checkmark \\
    \hline
    Tumblr & U.S. & \multicolumn{1}{m{2cm}}{2007} & A new media form between traditional blog and mocroblog. & & & & \checkmark \\
    \hline
    Instagram & U.S. & \multicolumn{1}{m{2cm}}{2010} &Share your pictures in a quick, wonderful and fun way.&&&\checkmark &\\
    \hline
    Pinterest & U.S. & \multicolumn{1}{m{2cm}}{2010} & A picture sharing social networking site. & & & \checkmark & \\
    \hline
    Line & Korea & \multicolumn{1}{m{2cm}}{2011} & Cross platform free communication software &\checkmark & & & \\
    \hline
    Snapchat & U.S. & \multicolumn{1}{m{2cm}}{2011} & A ``burn after reading'' photo sharing application & & &\checkmark & \\
    \hline
    Wechat & China & \multicolumn{1}{m{2cm}}{2011} & A free application providing instant messaging service for smart terminals. & \checkmark & & & \checkmark \\
    \hline
    Skype & U.S. & \multicolumn{1}{m{2cm}}{2013} & Global free voice communication software.&\checkmark & & & \\
    \hline
    Vine & U.S. & \multicolumn{1}{m{2cm}}{2013} & A social service system based on geographical location. & & & \checkmark & \\
    \hline
    Maimai & China & \multicolumn{1}{m{2cm}}{2013} & It is a real name career-oriented social platform that helps employees expand their contacts, communicate, cooperate, and apply for jobs. & & \checkmark & & \checkmark \\
    \hline
    DingTalk & China & \multicolumn{1}{m{2cm}}{2014} & A multi terminal platform for free communication and collaboration for Chinese Enterprises. & & \checkmark & &\\
    \hline
    WhatsApp & U.S. & \multicolumn{1}{m{2cm}}{2014} & An application for communication between smart phones & \checkmark & & & \\
    \hline
    \hline
    \end{tabular}
}
\end{table*}

The primary purpose of IoP is to connect people via Internet, so as to enable humans participate in various social activities. With the emergence of cyberspace, online social networks and social media have tried their best to meet the needs of human communications through a variety of means and tools. As a result, some social platforms come into being. As shown in Table \ref{Tab2}, we have listed some typical IoP-enabled platforms, and depicts their mainstream functions.

Based on our previous research on social media, we mainly divide their functions into four categories: instant messaging, career-oriented, multimedia research, and broadcast interaction. Instant messaging strengthens its function of supporting real-time communications, while career-oriented highlights its important role in career development. Multimedia search provides huge possibilities for users to search and scan information expected, and broadcast interaction allows each user to post your own news and interact with others. Note that the functions we describe here are the most common and popular instead of all of them.

\section{Open challenges and issues}

With the development of Internet and cyber techniques, IoP has vastly expanded into daily life and industrial manufacture. Meanwhile, it also brings about series of open issues and challenges needed to be discussed furthermore.

\begin{itemize}
\item Redundant identification: Since IoP focuses more on humans, identifying human nodes' problems is one of the most challenging issues. Humans may have complicated social roles and establish social relationships with different social networks, and there may exist various and redundant ways for identification. Therefore, guaranteeing the consistency and estimating the redundancy between different identifications and providing a recognized identification standard for human nodes should be researched in depth.
\item Resource optimization: This is about the resource optimization in a given IoP network. When it is confronted with limited communication or computing resources, it is so urgent to make a balance between various IoP nodes and maximize the efficiency. In other words, it is significant to own specific resource allocation mechanisms in given IoP networks.
\item Data sharing: As IoP is composed of various human nodes enabled by intelligent personal devices such as mobile phones, intelligent watches that may be different in data structures, types, and transmitting protocols. Therefore, there may exist barriers when data and information sharing. A standardized data sharing model should be initialized and updated in the future.
\item Service accuracy: As mentioned above, providing with humans the most appropriate services is a challenge that needs to overcome in IoP. In the following work, we still need to prompt technical innovations related to user portraits modeling and intelligent recommendation, so as to improve the accuracy of services.
\item Security and privacy: The security and privacy issues are so substantial since, in IoP, personal data and information of humans could be generated, transmitted, processed, and analyzed via the network. It is essential to research algorithms such as identity encryption, data encryption, protocol encryption, etc. to protect sensitive information as owners want.
\end{itemize}

\section{Conclusions}

In recent years the human-centric IoP has become one of the popular computing paradigms that overcomes physical boundaries and enables humans to be connected with tighter relationships. It deeply interconnects humans worldwide and allows to conduct with various social activities. In this paper, we give an overall survey of IoP, from its concept and enabling techniques. We also list and compare some common IoP-enabled platforms on the basis of different functions, envision open challenges and issues in IoP. The emergence of IoP is not to replace the original architecture of IoT, but an expansion and migration to social space with the emergence of spatial integration. It will explosively stimulate a new round of technological, ethical, and legal breakthroughs or reforms in the future.

\ifCLASSOPTIONcaptionsoff
  \newpage
\fi



\begin{thebibliography}{10}
\providecommand{\url}[1]{#1}
\csname url@samestyle\endcsname
\providecommand{\newblock}{\relax}
\providecommand{\bibinfo}[2]{#2}
\providecommand{\BIBentrySTDinterwordspacing}{\spaceskip=0pt\relax}
\providecommand{\BIBentryALTinterwordstretchfactor}{4}
\providecommand{\BIBentryALTinterwordspacing}{\spaceskip=\fontdimen2\font plus
\BIBentryALTinterwordstretchfactor\fontdimen3\font minus
  \fontdimen4\font\relax}
\providecommand{\BIBforeignlanguage}[2]{{%
\expandafter\ifx\csname l@#1\endcsname\relax
\typeout{** WARNING: IEEEtran.bst: No hyphenation pattern has been}%
\typeout{** loaded for the language `#1'. Using the pattern for}%
\typeout{** the default language instead.}%
\else
\language=\csname l@#1\endcsname
\fi
#2}}
\providecommand{\BIBdecl}{\relax}
\BIBdecl

\bibitem{A1}
L.~Atzori, A.~Iera, and G.~Morabito, ``The internet of things: A survey,''
  \emph{Computer networks}, vol.~54, no.~15, pp. 2787--2805, 2010.

\bibitem{A2}
H.~Ning, F.~Shi, S.~Cui, and M.~Daneshmand, ``From iot to future cyber-enabled
  internet of x (iox) and its fundamental issues,'' \emph{IEEE Internet of
  Things Journal}, 2020.

\bibitem{A3}
M.~Conti, A.~Passarella, and S.~K. Das, ``The internet of people (iop): A new
  wave in pervasive mobile computing,'' \emph{Pervasive and Mobile Computing},
  vol.~41, pp. 1--27, 2017.

\bibitem{A4}
D.~GL, ``Internet of people 2017 whitepaper,'' Available online:
  \url{https://www.dnv.com/life-sciences/internet-of-people/index.html}
  (accessed on 23 March 2021).

\bibitem{A5}
L.~Garton, C.~Haythornthwaite, and B.~Wellman, ``Studying online social
  networks,'' \emph{Journal of computer-mediated communication}, vol.~3, no.~1,
  p. JCMC313, 1997.

\bibitem{A6}
C.~Data and I.~Lab, ``The international conference on internet of people
  (iop2015),'' Available online:
  \url{http://www.cybermatics.org/SmartWorldCongress2015/SWC2015/IoP/IoP2015.htm}
  (accessed on 18 March 2021).

\bibitem{A7}
H.~Ning, \emph{Unit and ubiquitous internet of things}.\hskip 1em plus 0.5em
  minus 0.4em\relax CRC press, 2013.

\bibitem{A8}
J.~Johnson, ``Worldwide digital population as of january 2021,'' Available
  online:
  \url{https://www.statista.com/statistics/617136/digital-population-worldwide/}
  (accessed on 21 March 2021).

\bibitem{A9}
B.~Huang, M.~Chen, P.~Huang, and Y.~Xu, ``Gait modeling for human
  identification,'' in \emph{Proceedings 2007 IEEE International Conference on
  Robotics and Automation}.\hskip 1em plus 0.5em minus 0.4em\relax IEEE, 2007,
  pp. 4833--4838.

\bibitem{A10}
A.~Maedche and S.~Staab, ``Ontology learning for the semantic web,'' \emph{IEEE
  Intelligent systems}, vol.~16, no.~2, pp. 72--79, 2001.

\bibitem{A11}
E.~Raad, R.~Chbeir, and A.~Dipanda, ``User profile matching in social
  networks,'' in \emph{2010 13th International Conference on Network-Based
  Information Systems}.\hskip 1em plus 0.5em minus 0.4em\relax IEEE, 2010, pp.
  297--304.

\bibitem{A12}
H.~Ning, S.~Dhelim, and N.~Aung, ``Personet: Friend recommendation system based
  on big-five personality traits and hybrid filtering,'' \emph{IEEE
  Transactions on Computational Social Systems}, vol.~6, no.~3, pp. 394--402,
  2019.

\bibitem{A13}
Y.~Zou, J.~Zhu, X.~Wang, and L.~Hanzo, ``A survey on wireless security:
  Technical challenges, recent advances, and future trends,'' \emph{Proceedings
  of the IEEE}, vol. 104, no.~9, pp. 1727--1765, 2016.

\bibitem{A14}
WIKIPEDIA, ``Bluetooth,'' Available online:
  \url{https://en.wikipedia.org/wiki/Bluetooth} (accessed on 18 March 2021).

\bibitem{A17}
J.~Scott, ``Social network analysis,'' \emph{Sociology}, vol.~22, no.~1, pp.
  109--127, 1988.

\bibitem{A18}
P.~J. Carrington, J.~Scott, and S.~Wasserman, \emph{Models and methods in
  social network analysis}.\hskip 1em plus 0.5em minus 0.4em\relax Cambridge
  university press, 2005, vol.~28.

\bibitem{A19}
D.~Liben-Nowell and J.~Kleinberg, ``The link-prediction problem for social
  networks,'' \emph{Journal of the American society for information science and
  technology}, vol.~58, no.~7, pp. 1019--1031, 2007.

\bibitem{A20}
M.~Parameswaran and A.~B. Whinston, ``Social computing: An overview,''
  \emph{Communications of the Association for Information Systems}, vol.~19,
  no.~1, p.~37, 2007.

\bibitem{A21}
J.~Evans, ``Social computing unhinged,'' \emph{Journal of Social Computing},
  vol.~1, no.~1, p.~1, 2020.

\bibitem{A22}
L.-L. Shi, L.~Liu, Y.~Wu, L.~Jiang, M.~Kazim, H.~Ali, and J.~Panneerselvam,
  ``Human-centric cyber social computing model for hot-event detection and
  propagation,'' \emph{IEEE Transactions on Computational Social Systems},
  vol.~6, no.~5, pp. 1042--1050, 2019.

\bibitem{A23}
R.~Jin, H.-L. Zhang, and Y.~Zhang, ``The uncertainty problem in social
  computing and its solution method,'' in \emph{2018 International Conference
  on Robots \& Intelligent System (ICRIS)}.\hskip 1em plus 0.5em minus
  0.4em\relax IEEE, 2018, pp. 517--521.

\bibitem{A24}
K.-L. Skillen, L.~Chen, C.~D. Nugent, M.~P. Donnelly, W.~Burns, and I.~Solheim,
  ``Ontological user profile modeling for context-aware application
  personalization,'' in \emph{International conference on ubiquitous computing
  and ambient intelligence}.\hskip 1em plus 0.5em minus 0.4em\relax Springer,
  2012, pp. 261--268.

\bibitem{A25}
J.~Hu, F.~Jin, G.~Zhang, J.~Wang, and Y.~Yang, ``A user profile modeling method
  based on word2vec,'' in \emph{2017 IEEE International Conference on Software
  Quality, Reliability and Security Companion (QRS-C)}.\hskip 1em plus 0.5em
  minus 0.4em\relax IEEE, 2017, pp. 410--414.

\bibitem{A26}
G.~Fischer, ``User modeling in human--computer interaction,'' \emph{User
  modeling and user-adapted interaction}, vol.~11, no.~1, pp. 65--86, 2001.

\bibitem{A27}
T.~DuBois, J.~Golbeck, J.~Kleint, and A.~Srinivasan, ``Improving recommendation
  accuracy by clustering social networks with trust,'' \emph{Recommender
  Systems \& the Social Web}, vol. 532, pp. 1--8, 2009.

\bibitem{A28}
D.~F. Gurini, F.~Gasparetti, A.~Micarelli, and G.~Sansonetti, ``Temporal
  people-to-people recommendation on social networks with sentiment-based
  matrix factorization,'' \emph{Future Generation Computer Systems}, vol.~78,
  pp. 430--439, 2018.

\bibitem{A29}
S.~Dhelim, H.~Ning, and N.~Aung, ``Compath: User interest mining in
  heterogeneous signed social networks for internet of people,'' \emph{IEEE
  Internet of Things Journal}, 2020.

\end{thebibliography}

\end{document}